\documentclass[manuscript=article]{achemso} 
\usepackage[version=3]{mhchem} 
\usepackage{xcolor}
\usepackage{amsmath}
\usepackage{amssymb}
\usepackage{mathtools}
\usepackage[export]{adjustbox}
\usepackage{appendix}
\usepackage[labelformat=simple,labelsep=quad, skip=3pt]{subcaption}
\usepackage{graphicx}
\usepackage{url}

\usepackage[normalem]{ulem}
\usepackage[]{xcolor}

\author{Giuliano Migliorini}
\affiliation{Université d'Orléans, CNRS, 
                              CBM UPR 4301, Orléans, 45100, France}
\alsoaffiliation{Dipartimento di Fisica e Astronomia, Università degli studi di Firenze, Via Sansone 1, Sesto Fiorentino (FI), 50019, Italy}
\email{giuliano.migliorini@unifi.it}
\author{Josipa Cecic Vidos}
\affiliation{Université d'Orléans, CNRS, 
                              CBM UPR 4301, Orléans, 45100, France}
\author{Josef Hamacek}
\affiliation{Université d'Orléans, CNRS, 
                              CBM UPR 4301, Orléans, 45100, France}

\author{Anand Yethiraj}
\affiliation{Department of Physics, University of Guelph, Guelph, ON, Canada}     \email{ayethira@uoguelph.ca}
                         
\author{Francesco Piazza}
\affiliation{Dipartimento di Fisica e Astronomia, Università degli studi di Firenze, Via Sansone 1, Sesto Fiorentino (FI), 50019, Italy}
\alsoaffiliation{INFN sezione di Firenze, Via Sansone 1, Sesto Fiorentino (FI), 50019, Italy}
\email{francesco.piazza@unifi.it}

\title[An \textsf{achemso} demo]
{Physical Properties of
Dextran Solutions as Model Crowding Media} \color{black}
\abbreviations{NMR, PFG-NMR, FCS, SEC, MW}
\begin{document}
\footnotetext{G. Migliorini and J. Cecic Vidos contributed equally to this work and share first authorship. G. Migliorini analyzed most of the experimental data and developed their interpretation. J. Cecic Vidos performed all the experiments.}
\begin{tocentry}





\end{tocentry}

\begin{abstract}
\noindent The role of macromolecular crowding in living systems
is widely appreciated, but artificial crowders used to model these effects 
in vitro are often inadequately characterized. 
In this work, we examine density, viscosity, polymer self-diffusion 
and water diffusion in crowded dextran systems.
Dextran viscosity and self-diffusion follow size-dependent trends,
collectively described by universal functions of the overlap concentration
corresponding to a Flory exponent of 0.44, characteristic of branched polymers. 
Viscosity increases with concentration as a power law, with 
a crossover from dilute to semi-dilute behaviors. Dextran self-diffusion 
decays exponentially: this can be interpreted in light of Rosenfeld's excess 
entropy scaling hypothesis. 
Water self-diffusivity and specific volume decrease with concentration,  but
show no dependence on polymer size.
We show how these results can be used 
to construct the true volume fraction of crowders, 
which takes into account bound water.
Overall, our findings showcase the power of polymer physics concepts 
in macromolecular crowding studies in vitro.

\end{abstract}

\section{Introduction}
Real biological fluid media are crowded by macromolecules \cite{Alfano2024}. Concentrations of proteins on the order of 0.03 - 0.07 g/cm$^3$ are reported for interstitial fluids, 0.08 g/cm$^3$ for blood plasma\cite{zeiger2012macromolecular}, while the average intracellular protein concentration is estimated in the range 0.12-0.25 g/cm$^3$ for different human cells, together with an even higher figure of 0.3-0.4 g/cm$^3$ for the interior of \textit{E. Coli}\cite{Model2021}.  
These high concentrations profoundly impact the biochemical and physical properties of these media \cite{Minton2025}. However, molecular concentrations in commonly used buffers in cell culture and enzymology range only from $10^{-3}$ to $10^{-2}$ g/cm$^3$ [$\!$\citenum{zeiger2012macromolecular}] and do not accurately reflect the true complexity and conditions of the cellular or extracellular \textit{milieu}. For these reasons, the study of macromolecular crowding has attracted increasing interest within the scientific community \cite{Minton1980,ELLIS2001,Foffi2013, Matic2020,Alfano2024}. Macromolecular crowding increases the solution viscosity and hinders the self-diffusion of molecules \cite{mittal2015macromolecular}. Steric interactions with the macromolecular environment \cite{ZimmermanMinton1993} are expected to have an impact on the stability of native conformations of proteins \cite{Stagg2007, KimMittal2013} and hydration effects influence enzymatic activity \cite{VERMA2011}. 

Excluded volume is central to understanding the physics of macromolecular crowding 
\cite{Minton1980}. It is recognized that the volume excluded by a macromolecule to other solutes is a quantity that depends on the solute size \cite{ELLIS2001,Minton2001}. However, in both experimental and theoretical studies, macromolecular crowders are often approximated with compact objects of simple shape \cite{ZimmermanMinton1993}, which identifies excluded volume with no ambiguities. 

Living systems are mainly crowded by proteins but, in experimental studies, polymer crowders are also used very often, as they are considered inert with respect to the interactions with proteins \cite{Alfano2024} and because they are commercially available at low cost \cite{Speer2022}. Among polymer crowders, PEG (polyethylene glycol), Ficoll and dextran are the most common \cite{Speer2022}. While Ficoll is recognized as a cross-linked and branched sugar polymer \cite{venturoli2005ficoll, Ranganathan2022}, dextran is sometimes approximated as linear \cite{venturoli2005ficoll}. However, not all polymers are equal as crowders. PEG, for example, is known to become more hydrophobic with increasing temperature and molecular weight, and reportedly associates with proteins~\cite{Wu2013}. Ficolls were thought to be compact spherical polysaccharides, but recent work suggests they are far from compact~\cite{Ranganathan2022}. 

Structurally, dextran is, in fact, a branched polysaccharide. It has been studied extensively beyond its role as a macromolecular crowder, particularly in the context of medicine, pharmacology, and food industry\cite{MCCURDY1994, DiazMontes2021}. It is synthesized by some lactic acid bacteria and composed of a linear chain of D-glucose monomers, linked by $\alpha$(1-6) bonds, with branches linked by $\alpha$(1-2), $\alpha$(1-3) or $\alpha$(1-4) bonds \cite{DiazMontes2021}. Physical conditions, such as temperature and pH, the specific bacterial or enzymatic species involved in the synthesis, all together determine the branching properties of dextran \cite{DiazMontes2021}. Its degree of branching is estimated around 5\% for the very common dextran produced by \textit{Leuconostoc Mesenteroides} \cite{IoAbBu:Art}.
The Flory exponent for dextran, which describes the scaling of the polymer radius of gyration with its degree of polymerization, is not $0.5$ or $0.59$ (the values for a random or a real chain in a good solvent) but in the range $\nu$ $\sim$ 0.43 - 0.46: these are often attributed in the dextran literature to branching \cite{Nordmeier1993, venturoli2005ficoll, Antoniou2012}, a property that also characterizes other polysaccharides of glucose \cite{Bu:Art}. Some authors also investigated the topological features of dextran \cite{IoAbBu:Art,Nordmeier1993}. 
More specifically, Burchard and coworkers attempted to classify dextran either as
a hyperbranched or randomly branched polymer\cite{IoAbBu_3_2001}.

In this work, we characterize several dextran solutions by examining their viscosity using rheometry, and polymer and water diffusivity using pulsed-field-gradient (PFG)-NMR, with the aim of connecting the macromolecular crowding perspective with polymer physics. 
While scaling concepts from polymer physics predict power-law dependencies on molecular weight and concentrations spanning the dilute, semi-dilute, and concentrated regimes\cite{degennes1979scaling,doi1986theory}, these concepts do not yield explicit functional forms for expressing physical observables as continuous functions of concentration. This may be the reason why many other models, beyond the ones originating from the scaling concepts of polymer physics, have been introduced to describe diffusion in polymer media \cite{MASARO1999731}. 

While diffusion in polymeric systems might be expected to exhibit power-law scaling as a function of concentration, phenomenological observation of an exponential dependence as a function of concentration is reported in many different experimental systems~\cite{MASARO1999731, Phillies1986, PalitYethiraj2017, Ranganathan2022}. Our observations in this work are also consistent with this phenomenology, and we shed some light on this behavior, casting it in relation with excess-entropy scaling \cite{Rosenfeld1977,Dzugutov2001,Dyre2018}. 
In addition, we find (consistent with a previous report in Ficoll~\cite{Ranganathan2022}) that tracking the self-diffusivity of water molecules as a function of dextran concentration yields an estimate of the fraction of water molecules bound to dextran and consequently the total hydrodynamic volume of dextran.

Experimental measurements were carried out in two laboratories, one at the Centre Biophysique Mol\'eculaire (CBM) and the other at Memorial University of Newfoundland (MUN). 
\section{Materials and Methods}
\subsection{Materials}
Different sizes of dextran form \textit{Leuconostoc sp.} (9 to 11 kDa, 40 kDa and 450 to 600 kDa, further referred as DEX10, DEX40 and DEX450, see Table~\ref{tab:DextranMw}), deuterium oxide (D$_2$O),  and Dulbecco's Phosphate Buffered Saline (10x concentrated, PBS) were acquired from Sigma. 

\begin{table}[h!]
    \centering
    \begin{tabular}{c c c}
        \hline
        \textbf{Dextran sample} & $M_{\rm DEX}$ [g/mol]  & $N$ \\
         \hline
        DEX10   & $9.24 \times 10^3$ & 57\\
        DEX40   & $37.3 \times 10^3$ & 230\\
        DEX450  & $587 \times 10^3$ & 3620\\
        \hline
    \end{tabular}
    \caption{Weight average molecular weights $M_w$ of the samples declared by the manufacturer, and the corresponding degree of polymerization $N = M_{\rm DEX}/M_m$, where $M_m = 162$ g/mol is the monomer molecular weight. 
    }
    \label{tab:DextranMw}
\end{table}
\subsection*{Preparation of dextran solutions}
\noindent Stock solutions of dextran 10, 40 and 450 kDa were prepared by dissolving an appropriate amount of powder in 10x PBS and water (to have 1x final concentration). Solutions were then homogenised using the Fisherbrand 850 homogenizer (Fisher Scientiﬁc, USA) using a 7 mm × 115 mm dispersion element (11000 rpm, 7 cycles, each lasting 3 min with a 1 min break). Stock solutions were diluted with PBS to the intermediate concentrations used in the experiments to have final concentrations of 2.5, 5, 7.5, 10, 15 and 20 and 25\% (w/w).

\subsection{Density measurements}
\noindent To measure the density of dextran solutions, following their preparation as described in the previous paragraph, a METTLER-TOLEDO Density 30 px densitometer was used. This densitometer is not equipped with a thermostated chamber: thus, in order to measure the density at 37°C, solutions were pre-heated to 50°C and injected to the densitometer. The apparatus also measures the temperature of the solution, so the densities were recorded when the temperature was at 37°C. The experimental uncertainty related to the estimation of the solution density $\rho$ was propagated to mass concentrations $c$. This uncertainty in estimation of $c$  was taken into account in the fits of viscosity and diffusion.

\subsection{Viscosity measurements}
\noindent Rheology measurements were carried out using a 50-mm-diameter cone-plate (Anton Paar MCR 301) rheometer at MUN by measuring shear stress $\sigma$ (in Pa) for a range of shear rates $\dot{\gamma}$ (in $s^{-1}$). No other measurement modalities were employed because, in the range $1 < \dot{\gamma} < 600 \;s^{-1}$, all the dextrans probed showed a Newtonian (linear $\sigma$ vs. $\dot{\gamma}$) relation. The measurements were done at 37°C. The concentrations of each dextran used were: 5, 10, 15 and 20\% (w/w). 

\subsection{Water self-diffusion using PFG-NMR}
\noindent The diffusion of water in the presence of dextran was measured using the 400 MHz NMR instrument at CBM and applying a square pulse sequence. The sequence used was a variant of pulsed field gradient stimulated echo (ledbpgp2s). The solutions were prepared by diluting the stock solution (prepared as described before) of each dextran with PBS and D$_2$O to have final concentrations: 2.5, 5, 7.5, 10, 15 and 20 and 25\% (w/w) of each of the three dextran sizes and 10\% of D$_2$O. Due to H-D exchange, what is measured is the proton signal of HDO in water. We used as diffusion calibration the diffusion coefficient of trace HDO in D$_2$O at 298 K, $D = 1.902 \times 10^{-9} \mathrm{m^2/s}$.  All the samples were prepared in a volume of 250 µL and placed in 3 mm NMR tubes. Prior to each measurement, tubes were pre-incubated at 37°C inside of the machine as the measurements were taken at the same temperature. A sample of nearly pure D$_2$O (containing trace HDO) was measured under the same conditions each day before starting with the dextran samples and its diffusion coefficient was used as calibration. The signal intensity $S$ of the peak related to water molecules, as a function of the magnetic field gradient amplitude $g$, was fitted to a single exponential function, according to the Stejskal-Tanner equation \cite{Sinnaeve2012}, 
$S = S_0 \exp\left(-D \gamma^2\delta^2g^2 (\Delta - \delta/3) \right)$, where $\gamma$ is the proton magnetogyric ratio, $\delta$ is the duration of the gradient pulses, and $\Delta$ is the diffusion time. $S_0$ is the signal at zero applied gradient. A fit of the signal attenuation, with increasing $g$, to this equation yields the diffusion coefficient $D$. 

\subsection{Dextran self-diffusion measurements using PFG-NMR}
\noindent Samples for PFG-NMR were prepared by mixing different quantities of dextran solution with 10:90 D$_2$O:water (i.e., 90\% by weight of distilled deionized water mixed with 10\% by weight of D$_2$O). The final dextran concentration ranged from 2.5 up to 25\%. 
The tubes used were 3 mm NMR tubes and all measurements were carried out at 37°C. Each day a sample of D$_2$O at 37°C was measured prior to measuring the dextran samples as a calibration. Measurements were executed using a Bruker Avance II NMR
spectrometer at a 1H resonant frequency of 600 MHz at MUN equipped with a diffusion probe with a maximum field gradient of 1800 G/cm. The pulse program used was pulsed field gradient stimulated echo (diffSte).

The average contribution to dextran self diffusion was estimated considering only the peak intensity related to the four lower available values of the magnetic gradient for each crowding condition, in the range 5 - 500 G/cm. In this regime, peak intensity was fitted to a single exponential function of gradient, according to the Stejskal-Tanner equation \cite{Sinnaeve2012}, and the self-diffusion coefficient was extracted. Contributions to self-diffusion at higher gradients were not considered for this publication, although the trends deviated from a single exponential decay. It has been noted previously that this early decay is consistent with the {\textit mean} diffusion coefficient when there are multiple diffusivities due either to aggregates or polydispersity.~\cite{Barhoum2016a}
The values of dextran self-diffusion reported in the results are carefully cross-calibrated against the $D_2O$ daily measurement to ensure consistency of measurements across different days.

\section{Results}

\subsection{Density measurements yield dextran specific volumes}

\begin{figure}[h!]
    \centering
    \includegraphics[width=12 cm]{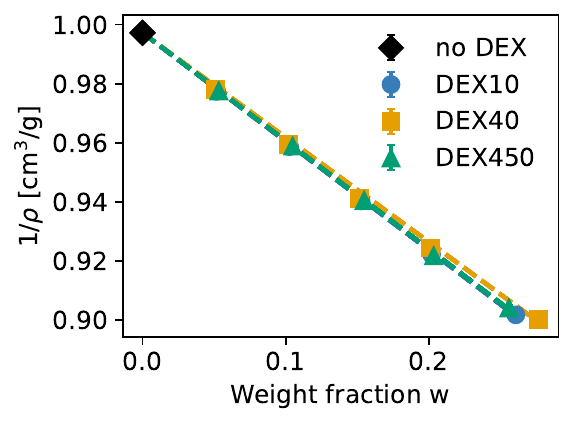}
    \caption{Specific volume of the solution, $1/\rho(w)$ shows a linear decrease as a function of dextran weight fraction $w$. Dashed lines correspond to fits according to Eq.~\eqref{eq:density_w} with specific volumes $v_0$ = $1/\rho_0$ = $(0.9972 \pm 0.0011)$ cm$^3$/g and $\overline{v}$ reported in Table \ref{tab:Partialspecvolumes}.}
    \label{fig:density}
\end{figure}

Experimental measured solution densities (as described in Methods) are shown in Figure~\ref{fig:density}: the inverse density $1/\rho$, which is the specific volume of the dextran solution, was plotted as a function of the solute weight fraction $w$. The results were fitted to the expected relation
\begin{equation}
    \label{eq:density_w}
    \frac{1}{\rho (w)} = v_0 (1 - w) + \overline{v} w,
\end{equation} 
where $v_0$ and $\overline{v}$ are the solvent and solute specific volumes, respectively. Such an approach assumes that both these quantities can be considered constant in the explored experimental range. $v_0$ is determined experimentally as $v_0$ = $1/\rho_0$ = $(0.9972 \pm 0.0011)$ cm$^3$/g. 
Good agreement of the fit with the data in Figure \ref{fig:density} is observed. The specific volume of all three dextrans is estimated around $\overline{v}$ $\simeq$ 0.63 cm$^3$/g, with no significant differences between the three (see Table \ref{tab:Partialspecvolumes}).
%
\begin{table}[ht!]
    \centering
    \begin{tabular}{r c}
        \hline
        \textbf{Mass [kDa]} & $\overline{v}$ [cm$^3$/g] \\
         \hline
        10   & $0.629  \pm  0.001$ \\
        40   & $0.643  \pm  0.003$ \\
        450  & $0.631  \pm  0.001$ \\
        \hline
    \end{tabular}
    \caption{Partial specific volumes for different average molecular weight dextran.}
    \label{tab:Partialspecvolumes}
\end{table}
%
From the estimation of $\rho(w)$ and weight fractions $w$, 
the mass concentrations were calculated as $c = w\rho(w)$.
In the following, we will use $c$ as the variable. 
It can be easily verified that, substituting $w = c/\rho(c)$ 
in eq. \ref{eq:density_w}, the density of the solution $\rho(c)$ is linear 
if expressed as a function of the mass concentration $c$
\begin{equation}
    \label{eq:density_c}
    \rho(c) = \rho_0 + c\left(1 - \overline{v}\rho_0\right).
\end{equation}
%

\subsection{Dextran viscosities show universal concentration-dependent scaling}
\begin{figure}[htbp]
\centering
\begin{subfigure}{0.48\textwidth}
    \centering
    \includegraphics[width=\textwidth]{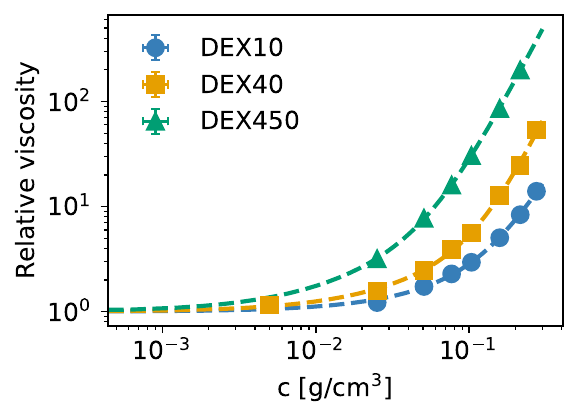}
    \caption{}
    \label{fig:vis-a}
\end{subfigure}
\hfill
\begin{subfigure}{0.48\textwidth}
    \centering
    \includegraphics[width=\textwidth]{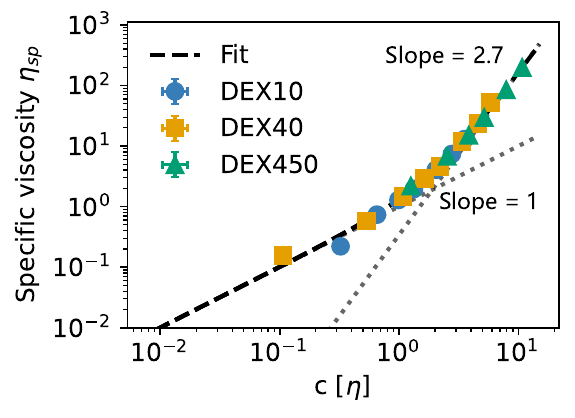}
    \caption{}
    \label{fig:vis-b}
\end{subfigure}
\caption{
\subref{fig:vis-a} 
Relative viscosity $\eta_{rel}$ of dextran solution as a function of concentration.
Dashed lines correspond to the best fits according to Eq.~\eqref{eq:vis_plaw}, with the parameters reported in Table~\ref{tab:rel-vis}.
\subref{fig:vis-b}
Specific viscosity $\eta_{sp} = \eta_{rel} - 1$, plotted as a function of the scaled (and dimensionless) variable $c \, [\eta]$, shows collapse of data from all molecular weights on a single curve. A crossover from a slope of unity (dilute solution behavior) to a slope $x = 2.7 \pm 0.1$ in the semi-dilute regime is seen in all cases.
\label{fig:vis}}
\end{figure}

The viscosity of dextrans (DEX10, DEX40 and DEX450) is plotted in 
Figure~\ref{fig:vis-a} as a function of mass concentration $c$.
It is useful to define a relative viscosity $\eta_{rel}$ and a specific viscosity $\eta_{sp}$ as follows:
\begin{align}
\label{eq:vis_plaw}
\eta_{rel} = \frac{\eta}{\eta_0} = 1 + [\eta]c + k_H ([\eta]c)^x, \\ \nonumber
\eta_{sp} = \eta_{rel} - 1 = [\eta]c + k_H ([\eta]c)^x.
\end{align}
A consistent increase in $\eta_{rel}$ is seen for all Dextran molecular weights, and Eq.~\ref{eq:vis_plaw} provides an excellent fit to all the results in Figure~\ref{fig:vis-a}.
Eq.~\eqref{eq:vis_plaw} is a phenomenological function~\cite{tuinier1999concentration, DeGennes1976a, Ranganathan2022} that is designed to capture the crossover from dilute solution behaviour (the first, linear $c$-dependent term) to the concentrated de Gennes reptation regime (the second, power law term).  $\eta_0$ is the solvent viscosity, measured to be $(0.702 \pm 0.006)\times 10^{-3}$ Pa s. The fit captures the transition between the two regimes and its results are reported in Table \ref{tab:rel-vis}. 
\begin{table}[ht!]
    \centering
    \begin{tabular}{c c c c}
        \textbf{Sample} & Intrinsic viscosity $[\eta]$ [cm$^3/$g] & $ k_H $ & $x$ \\
        \hline\hline
        DEX10  & $11.7 \pm 1.3$ & $0.4 \pm 0.2$ & $2.6 \pm 0.2$ \\
        DEX40  &  $25 \pm 2$ & $0.11 \pm 0.06$ & $3.1 \pm 0.2$  \\
        DEX450 & $73 \pm 3$ & $0.071 \pm 0.014$ & $2.85 \pm 0.04$ \\
        \hline
    \end{tabular}
        
    \caption{Intrinsic viscosity $[\eta]$, Huggins constant $k_H$ and power 
    law exponent $x$ for dextran of varying molecular weights when fitting the data individually through Eq.~\eqref{eq:vis_plaw}.}
    \label{tab:rel-vis}
\end{table}
The exponents $x$, when each dataset is fitted separately, are all in the range 2.7 - 3.1, while the best estimation of the Huggins constant $k_H$ appears to decrease with increasing size. Although the quality of the fit is good, one would not expect $x$ nor $k_H$ to depend on dextran size; moreover, the differences in their values is not significant considering the uncertainties (reported in Table \ref{tab:rel-vis}). Only the variation with molecular weight of the intrinsic viscosity is meaningful.

\subsubsection{Cumulative fitting of viscosity}
We carry out a cumulative fit of the relative viscosity results through equation \ref{eq:vis_plaw}, constraining the parameters $k_H$ and $x$ to be the same for all molecular weights, but constraining $[\eta]$ to depend on the dextran degree of polymerization $N$ through a power law that is inspired by the Mark-Houwink equation~\cite{RubinsteinColby} and is common in viscosimetry \cite{kulicke_viscosimetry_2004}:
\begin{equation}
    \label{eq:Mark-Houwink}
    [\eta] = K_{[\eta]} {N}^a.
\end{equation}
Here, $N$ is calculated from the polymer weight-averaged molecular weight $M_{\rm DEX}$ (reported by the manufacturer), i.e., $N = M_{\rm DEX}/M_m$, where $M_m = 162$ g/mol is the molecular weight of the monomer of dextran $C_6H_{10}O_5$, which is a glucose molecule deprived of two hydrogen and one oxygen atoms. Both $N$ and $M_{\rm DEX}$ for the three dextrans are reported in Table~\ref{tab:DextranMw}. 
\noindent We can now fit the relative viscosities in Figure~\ref{fig:vis-a} replacing $[\eta]$ with equation 4, so that $K_{[\eta]}$ and $a$ are free parameters that also have the same value for all three polymer molecular weights. The results of the fit are reported in Table \ref{tab:MHfitparams}.
\begin{table}[h!]
    \centering
    \begin{tabular}{rrr}
        {\bf Parameter} & best-fit value & Units \\
        \hline\hline
        $k_H$        &   $0.34 \pm 0.15$    & $-$       \\
        $K_{[\eta]}$ &   $3.6 \pm 0.4$      & cm$^3$/g  \\
        $a$          &   $0.321 \pm 0.007$  & $-$       \\
        $x$          &   $2.7 \pm 0.1$      & $-$       \\
        \hline
    \end{tabular}
    \caption{Fit parameters from the cumulative fit of relative viscosity results to the Huggins equation \ref{eq:vis_plaw}, expressing intrinsic viscosity $[\eta]$ using equation~\ref{eq:Mark-Houwink}.}
    \label{tab:MHfitparams}
\end{table}

In this approach, the intrinsic viscosity $[\eta]$ can be estimated according to its definition (eq. \ref{eq:Mark-Houwink}) for the three dextran sizes: $[\eta]_\mathrm{DEX10}\simeq 13.1$ cm$^3$/g; $[\eta]_\mathrm{DEX40} \simeq 20.5$ cm$^3$/g; $[\eta]_\mathrm{DEX450} \simeq 49.8$ cm$^3$/g. These values are close to previous estimates~\cite{Guner1999,antoniou2010structure}, at 37 and 40$\;$°C, on solutions of dextrans of similar size, and can be used to estimate the overlap  concentration $c^*$, which constitutes an upper limit to the dilute regime. 
The overlap concentration is defined as \cite{doi1986theory}
\begin{equation}
    \label{cond:c_overlap}
    \left(\frac{c^*N_A}{NM_m}\right)\left(\frac{4}{3}\pi {R}^3\right) \simeq 1
\end{equation}
where $N_A$ is the Avogadro number and $R$ is the polymer radius of gyration. 
In the regime $c > c^*$ the inter-polymer interaction cannot be neglected and the solutions are called semi-dilute, distinguishing them from the concentrated regime, achieved for $c \gg c^*$. 

The exponent $a$ provides an estimate of the Flory exponent $\nu$ for dextran through the relation $ a = 3\nu - 1 $, because $c^*$ scales~\cite{DeGennes1976a} with $N$ as $N^{1-3\nu}$, obtaining $\nu = 0.443 \pm 0.003$. 
Values of $\nu \sim 0.44$ as the scaling exponent for the radius of gyration of dextran are quite common in the literature and attributed to its branched nature \cite{Bu:Art,Nordmeier1993,Antoniou2012}. Moreover, a consistent value of $0.33$ for $a$ had also been previously reported studying the rheology of dextran solutions, with the low value of $a$ attributed to the branched nature of dextran samples \cite{TIRTAATMADJA2001295} . 

Scaling concepts from polymer physics predict that the relative viscosity is an invariant function of $c/c^*$ \cite{doi1986theory}. Therefore, the inverse of the intrinsic viscosity obtained through the cumulative fit can indicate the trend for the overlap concentration $c^*_\eta \sim 1/[\eta]$ \cite{IoAbBu:Art}. We demonstrate the quality of the cumulative fit by plotting the specific viscosity $\eta_{sp} \equiv \eta_{rel} - 1$ as a function of $c \, [\eta]$ in Figure~\ref{fig:vis-b}, which shows complete collapse of the results from the three molecular weights. We see that the asymptotes for the dilute and semi-dilute scaling behaviors intersect at $c^*_\eta [\eta] \sim 2$, suggesting a relationship $c^*_\eta = 2/ [\eta]$, and yielding $c^*_\mathrm{DEX10}\sim 0.152$ g/cm$^3$; $c^*_\mathrm{DEX40}\sim 0.098$ g/cm$^3$; $c^*_\mathrm{DEX450}\sim 0.040$ g/cm$^3$.
The power law scaling of the specific viscosity in the semi-dilute regime 
($c \, [\eta] > 2$) in Figure \ref{fig:vis-b}, is found to be $x = 2.7 \pm 0.1$. This value is much lower than the slope of $4.9$ reported by Ioan et al. \cite{IoAbBu:Art}. However, in that work the authors were able to explore higher values of $c \,[\eta]$, reaching values around $30$, using higher molecular weight dextran samples than those included in this assay. Therefore, it might be hypothesized that the power law behavior that is found in this paper underestimates the value because it lacks statistics at high enough $c \,[\eta]$. 
The range of values estimated for the Huggins constant, $k_H = 0.34 \pm 0.15$, overlaps with the range 0.42-0.61 determined by Antoniou and Tsianou on dextran samples of different molecular weights in water at 40°C \cite{Antoniou2012}. 

\subsection{Dextran diffusivities show exponential concentration dependence}
\begin{figure}[htbp]
\centering

\begin{subfigure}{0.47\textwidth}
    \centering
    \includegraphics[width=\textwidth]{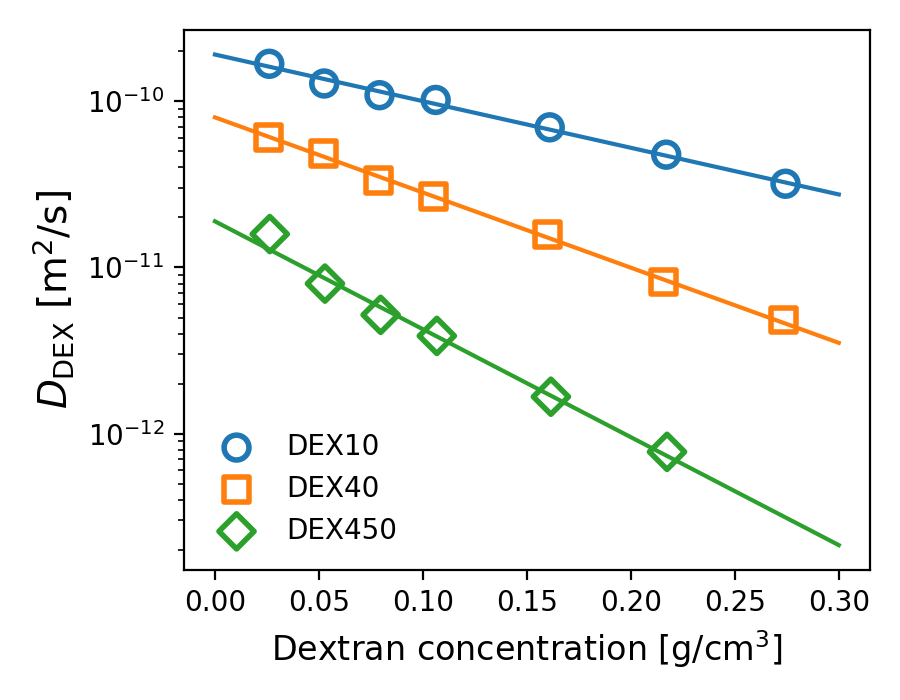}
    \caption{}
    \label{fig:DDEX-a}
\end{subfigure}
\hfill
\begin{subfigure}{0.5\textwidth}
    \centering
    \includegraphics[width=\textwidth]{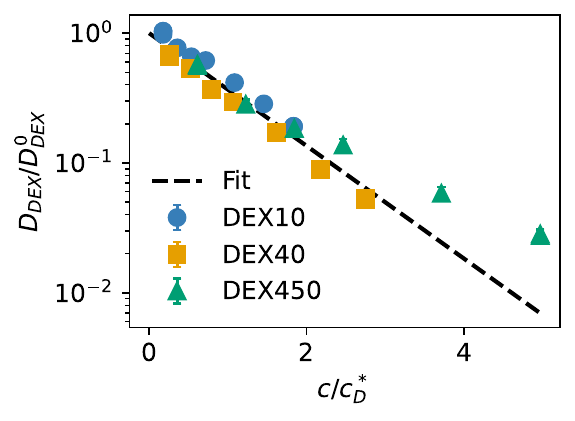}
    \caption{}
    \label{fig:DDEX-b}
\end{subfigure}
\caption{
\subref{fig:DDEX-a}
Self-diffusion coefficient of dextran. PFG-NMR measurements (symbols) and fits (solid lines) performed with Eq.~\eqref{e:DDEXfit}. The best-fit values of the fitting parameters are listed in Table~\ref{tab:diffusionweights_mc}.
\subref{fig:DDEX-b}
Reduced self-diffusion coefficient of dextran $D_{\rm DEX}/D_{\rm DEX}^0$, scaled by the concentration $c^*_D$, according to equation \eqref{eq:scaleddiff}, with $D_0 = d_1N^{-\nu}$ and $c^*_D = c_1N^{1 - 3\nu}$:  DEX10 (blue circles), DEX40 (orange squares), DEX450 (green triangles). The best fit function is reported as a black dashed line. The deviation from the fit for $c/c^*_D > 2$ in the case of DEX450 is possibly a consequence of approaching the semi-dilute regime.
}
\label{fig:DDEX}
\end{figure}

Dextran solutions were studied using PFG-NMR. 
The dextran self-diffusion coefficient, $D_\mathrm{DEX}$, shows a sharp decrease with concentration: this decrease was well fit to the relation
\begin{equation}
    \label{e:DDEXfit}
    D_{\rm DEX}(c) = D_{\rm DEX}^0 \, e^{-c/c^*_{D}}
\end{equation}
obtaining the results reported in Fig.~\ref{fig:DDEX-a}, where the $y$ axis is on displayed on a log scale to show clearly any deviations from exponential dependence as a deviation from linearity. The fit parameters,  $D_{\rm DEX}^0$ (diffusion coefficient at infinite dilution) and $c^*_{D}$ are tabulated in Table~\ref{tab:diffusionweights_mc}: while the latter is not necessarily the overlap concentration, it is interesting to note that it tracks the trend of the overlap concentration $c^*_{\eta}$, obtained from viscosity measurements (rightmost column), very well.

%

%

\begin{table}[h!]
    \centering
    \begin{tabular}{c c c c}
        \textbf{Sample} & $D_{\rm DEX}^0 \ [10^{-10}$ m$^2/$s] & $c^*_{D}$ [g/cm$^3$] & $c^*_{\eta}$ [g/cm$^3$]\\
        \hline\hline
        DEX10  &  $ 1.91 \pm 0.04 $  &  $0.155 \pm 0.003$ &  $0.152 \pm 0.017$\\
        DEX40  &  $ 0.80 \pm 0.01 $  &  $0.096 \pm 0.001$ &  $0.098 \pm 0.007$\\
        DEX450 &  $ 0.18 \pm 0.01 $  &  $0.068 \pm 0.003$ &  $0.040 \pm 0.002$\\
        \hline
    \end{tabular}
    \caption{Fitted self-diffusion coefficients $D_{\rm DEX}^0$ and characteristic concentrations $c^*_{D}$ (see eq.~\ref{e:DDEXfit})
    for dextrans of varying molecular weights. The overlap concentration obtained from visoscity measurements $c^*_{\eta}$ is shown in the last column for comparison.} 
    \label{tab:diffusionweights_mc}
\end{table}
%
It should be noted that alternatives to an exponential fit are less satisfactory: 
power-law forms, though theoretically motivated by scaling arguments in polymer physics\cite{degennes1979scaling,Teraoka2002}, require additional fitting parameters without providing a significant improvement in quality. The absence of a clear power-law regime may simply reflect the fact that our measurements do not extend to sufficiently high concentrations to probe the crossover toward the concentrated regime. Moreover, exponential trends for the diffusion as a function of concentration are  common in the literature \cite{MASARO1999731,PalitYethiraj2017}, either due to a phenomenological approach \cite{DOSTER20071360} or to connect diffusion with thermodynamic quantities \cite{PuellesHoyuelos2024}.
\subsubsection{Cumulative fitting of dextran self-diffusion}
\begin{figure}[ht!]
  \includegraphics[width=10.cm]{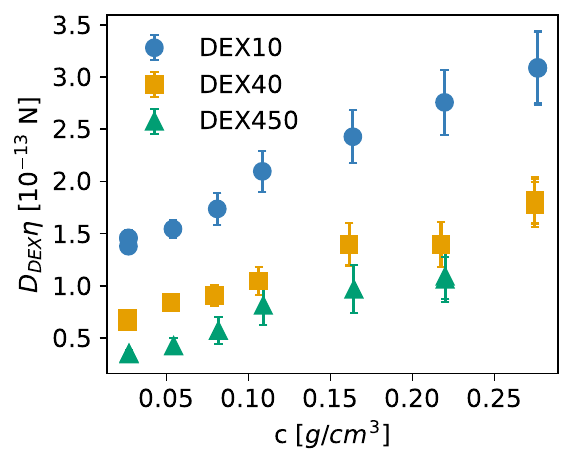}
   \caption{Product of dextran self-diffusion $D$ and viscosity $\eta$ of dextran solutions. $D\eta$ is reported in Newton (N).  At small concentrations, the difference in the $D_{\rm DEX}^0$ constant for the different sizes is the leading contribution to $D_{\rm DEX}.\eta$ at the lower concentrations. However the viscosity increases faster with concentration and determines the monotonic increase of $D\eta$. }
   \label{fig:Deta}
\end{figure}

In analogy with the data collapse observed for the viscosity, we attempt a cumulative fit of the data for all dextran molecular weights. 
In dilute conditions, the diffusion constant $D_{\rm DEX}^0$ is expected to scale with the inverse of the radius of the polymer $R$, as expected by the Stokes-Einstein relation: $D_{\rm DEX}^0 \sim R^{-1}$.
Since $R \sim N^\nu$, we can write $D_{\rm DEX}^0 = d_1 N^{-\nu}$, where $N$ is the degree of polymerization defined earlier. The scaling concentration $c^*_D$, instead, is expected to scale like the overlap concentration: $c^*_D \sim N/R^{3} \sim N^{1-3\nu}$. Taking $c^*_D = c_1 N^{1-3\nu}$, we introduce the following cumulative function to fit all the data simultaneously
\begin{equation}
    \label{eq:scaleddiff}
    D_{\rm DEX} = d_1 N^{-\nu}e^{-\frac{c}{c_1}N^{3\nu-1}}
\end{equation}
with three parameters: $d_1$, $c_1$ and the Flory exponent $\nu$. For the degree of polymerization $N$, we use the values calculated in Table~\ref{tab:DextranMw}. From the fits, the best estimate of the fitting parameters is: $d_1$ = $\left(1.01 \pm 0.07\right)\times 10^{-9}$ m$^2$/s, $c_1$ = $\left(0.60 \pm 0.09\right)$ g/cm$^3$, and $\nu = 0.444 \pm 0.009$. The value obtained for $\nu$ is in perfect agreement not only with the estimation obtained from viscosity in this work, but also with previous estimations of the scaling of dextran  self-diffusion in dilute conditions. Extrapolating the infinite dilution value from a dynamic light scattering experiment, Nordmeier estimated $\nu$ in the range $0.438 - 0.446$ varying the temperature from 15 to 60 °C in water \cite{Nordmeier1993}, while Maina, Pitkanen and collaborators found $\nu =  0.425$ with a DOSY PFG-NMR experiment in $D_2O$ \cite{MAINA2014199}. 
The result of the fit captures the self-diffusion of the DEX10 and DEX40 dextrans very well, as shown in Figure~\ref{fig:DDEX-b}, in the whole investigated concentration range. For DEX450, the collapse is equally good at low concentrations, but deviates when $c/c^*_D > 2$. 
It is notable that the exponential collapse occurs in the dilute-polymer regime, while the deviation occurs at concentrations that correspond to the transition to the semi-dilute regime.
If a generalized Stokes-Einstein relation (i.e., treating the concentration-dependent viscous background as an effective medium rather than as a crowded solution) were to be valid, $D_{\rm DEX}.\eta$ would be constant, for each dextran size. However, we instead see in Figure \ref{fig:Deta} that the product $D_{\rm DEX}. \eta$ increases with concentration, i.e., $\eta$ increases far more rapidly than $D_{\rm DEX}$ decreases. This is consistent with observations in Ficoll solutions~\cite{rashid2015macromolecular,palit2017effect} that a microscale measure of viscosity ($via$ diffusivity) increases more modestly than the bulk viscosity.

\subsection{Water diffusion in dextran solutions yields the bound water fraction}
\label{subs:water_diff}

\begin{figure}[htbp]
\centering

\begin{subfigure}{0.48\textwidth}
    \centering
    \includegraphics[width=\textwidth]{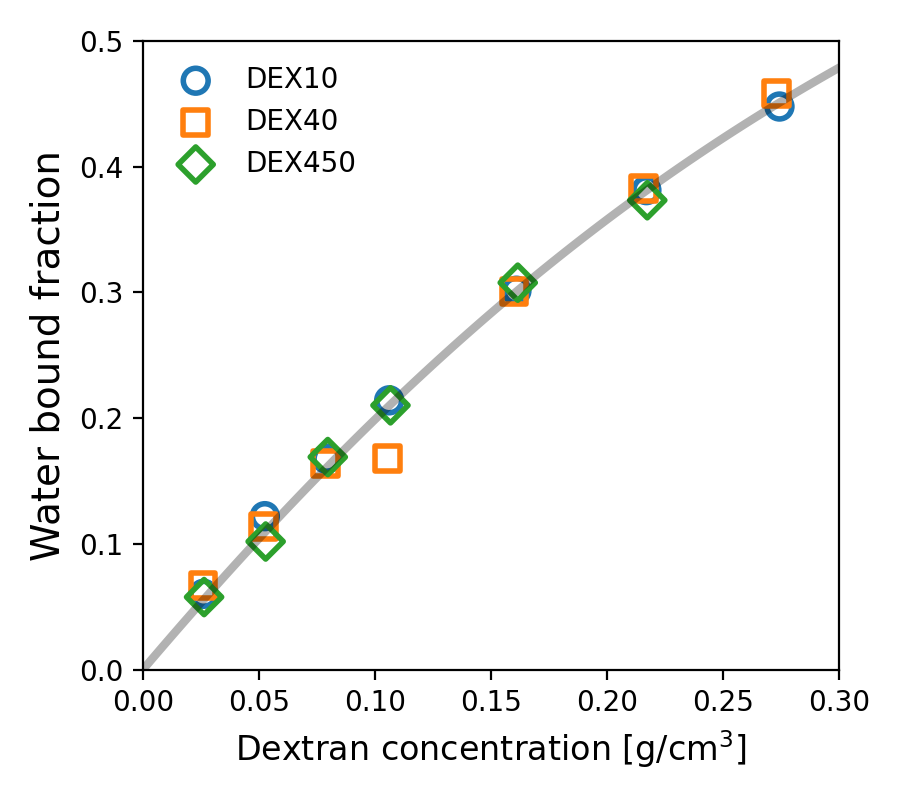}
    \caption{}
    \label{fig:fcDw-a}
\end{subfigure}
\hfill
\begin{subfigure}{0.48\textwidth}
    \centering
    \includegraphics[width=\textwidth]{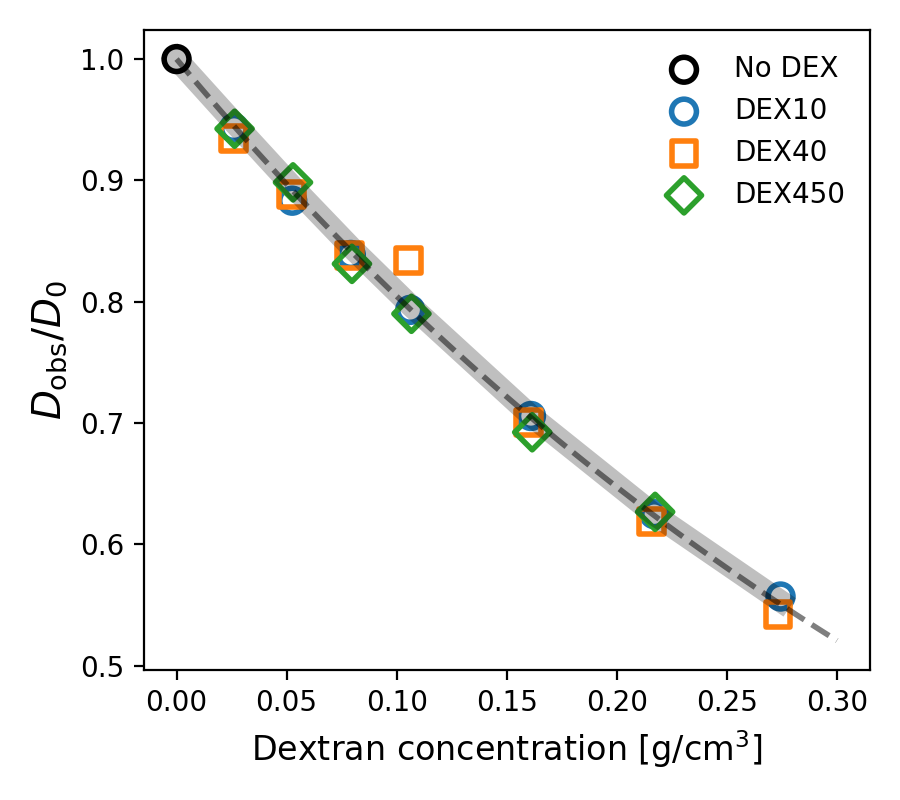}
    \caption{}
    \label{fig:fcDw-b}
\end{subfigure}
\caption{
\subref{fig:fcDw-a}
Symbols: Bound fraction of water molecules computed from the PFG-NMR measurements 
of water and dextran diffusion from Eq.~\eqref{eq:fcexp}.
Solid line: fit with expression~\eqref{eq:f_solution}. Best-fit 
parameters: $r= 36.995 \pm 0.001$, $K_d=50.52\pm0.03$ M. 
\subref{fig:fcDw-b}
Reduced diffusion coefficient of water vs dextran concentration. Direct PFG-NMR 
measurements (symbols) and best fit obtained by combining Eq.~\eqref{eq:DH20}
with the best-fit approximation of $f(c)$ through 
Eq.~\eqref{eq:f_solution}  (thick light gray line). 
The dashed line is a plot of the simple exponential approximation to the true 
solution, Eq.~\eqref{eq:Dnexp}, with the predicted value of the 
average number of contacts per monomer, $N_b^0 = 19.5$.
Other parameters are: $\rho_0$ = 1.00 g/cm$^3$, $\overline{v}$ = 0.63 cm$^3$/g.
}
\label{fig:fcDw}
\end{figure}

\noindent PFG-NMR measurements of water diffusion were first performed  
in dextran-free solutions (10\% D$_2$O in H$_2$O with PBS buffer).
This gave us the reference value $D_0 = \left(2.93 \pm 0.03\right) \times 10^{-9}$ $m^2/s$ at 37°C~\cite{Holetal}. 
In the presence of dextran polymers, water molecules will diffuse from the bulk to the highly hydrophilic 
dextran matrix and vice versa. If the exchange rates are reasonably rapid
(or equivalently, the residence times in either phase sufficiently short),
the water diffusion coefficient observed from the PFG-NMR decay, $D_{\rm obs}(c)$, will be 
a linear combination of the free water diffusion, $D_0$, and the bound 
water diffusion~\cite{Ranganathan2022}, 
which is essentially identical to the diffusion of dextran, $D_{\rm DEX}(c)$. Note that the observed and bulk water diffusion coefficients are indicated as $D_{\rm obs}$ and $D_0$ respectively. 
If we denote the equilibrium fraction of bound water at dextran 
concentration $c$ as $f(c)$, we should expect 
\begin{equation}
\label{eq:DH20}
D_{\rm obs}(c) = (1-f(c)) D_0 + f(c) D_{\rm DEX}(c)
\end{equation}
Since we have measured $D_{\rm DEX}(c)$ independently, 
the measured values of water and dextran diffusion can be combined to 
compute an experimental measurement of the bound fraction $f(c)$. 
Rearranging Eq.~\eqref{eq:DH20}, we have 
\begin{equation}
    \label{eq:fcexp}
    f(c) = \frac{D_0-D_{\rm obs}(c)}{D_0-D_{\rm DEX}(c)}
\end{equation}
These experimental estimates can be compared to the prediction of a simple 
equilibrium two-state model, 
providing at the same time a simple theoretical framework for water diffusion 
in a crowded medium. Let us imagine that free water molecules 
in the bulk, $W_f$, can bind to free binding sites, $R_f$, on dextran monomers, 
with a binding constant $k_{\rm on}$, forming a complex $R_b$, 
which dissociates with rate constant $k_{\rm off}$,
\begin{equation}
    \label{eq:bind_model}
    \ce{W_{\textit{f}} + R_{\textit{f}} <=>[k_{\rm on}][k_{\rm off}] R_{\textit{b}}} 
\end{equation}
Accordingly, the equilibrium (dissociation) constant is
$K_d = k_{\rm off}/k_{\rm on} = [W_f][R_f]/[R_b]$.
The fraction of bound water molecules is defined as 
\begin{equation}
    \label{e:fcdef}
    f = \frac{[R_b]}{[W_f] + [R_b]},
\end{equation}
In our system there are two conservation laws, one for water molecules 
and one for binding sites. On the one hand, 
letting $r$ denote the number of binding sites per dextran monomer, the 
following conservation law for the total number of binding sites should hold
at any dextran concentration $c$,
\begin{equation}
    \label{eq:R_f+R_b}
    [R_f] + [R_b] = \frac{rc}{M_m}    
\end{equation}
where $M_m$ = 162 g/mol is the molar mass of a dextran monomer.
On the other hand, a similar conservation law should also hold for the total number of water 
molecules
\begin{equation}
    \label{eq:W_f+R_b}
    [W_f] + [R_b] = \frac{\rho(c) - c}{M_w}
\end{equation}
where $M_w$ = 18.01 g/mol is the molar mass of water and we recall that 
$\rho(c)$ denotes the overall density of the dextran solutions, and $c$ the dextran mass concentration. From the definition~\eqref{e:fcdef} it is easy to see that
\begin{equation}
    \label{eq:Kd2}
    f K_d = (1 - f)[R_f]
\end{equation}
Recalling the two conservation laws~\eqref{eq:R_f+R_b} 
and~\eqref{eq:W_f+R_b}, we immediately obtain from this
\begin{equation}
    \label{eq:R_f}
    [R_f] = \frac{rc}{M_m} - f \left(\frac{\rho(c) - c}{M_w}\right)
\end{equation}
Finally, plugging Eq.~\eqref{eq:R_f} in Eq.~\eqref{eq:Kd2}, we obtain
a second order equation for the  bound fraction $f$,
\begin{equation}
    \label{eq:f}
    f^2 - \left( 1 + P(c) + Q(c) \right) f + Q(c) = 0
\end{equation}
where
\begin{equation}
    P(c) = \frac{K_d M_w}{\rho(c) - c} \qquad
    Q(c) = \frac{r c M_w}{M_m\left(\rho(c) - c\right)}
\end{equation}
The sought-for bound fraction is the root of Eq.~\eqref{eq:f}
that satisfies the condition $f(0)=0$, namely
\begin{equation}
    \label{eq:f_solution}
    f(c) = \frac{1 + P(c) + Q(c)}{2}\left(1 - \sqrt{1 - \frac{4Q(c)}{\left(1 + P(c) + Q(c)\right)^2}}\,\right). 
\end{equation}
Fig.~\ref{fig:fcDw}\subref{fig:fcDw-a} shows the experimental estimates 
of the water bound fraction obtained from Eq.~\eqref{eq:fcexp}. Interestingly, 
water dynamics in our dextran solutions does not depend on the size of the polymers.
Rather, it is seen to follow a universal trend for all molecular weights of Dextran. Also shown is the fit with our 
simple model that gives $r= 36.995 \pm 0.001$ and $K_d= (50.52\pm 0.03)$ M. 
The model curve $f(c)$ can be plugged into Eq.~\eqref{eq:DH20} and the resulting $D_{\rm obs}(c)/D_0$ compared directly 
to the PFG-NMR measurements of water diffusion in Fig.~\ref{fig:fcDw}\subref{fig:fcDw-b}. Thus, our model provides a quantitative rationale for the apparent observed reduction of water diffusivity 
in increasingly crowded solutions. 
%

%

\indent The model also affords more transparent insight into the binding process of water molecules 
in the dextran matrix. Recalling Eq~\eqref{eq:R_f+R_b},
the equilibrium number of bound water molecules $N_b$ per monomer can be calculated as
\begin{equation}
    \label{eq:N_b}
    N_b = \frac{M_m [R_b]}{c} = 
          \frac{M_m f(c) \left(\rho(c) - c\right)}{c \,M_w}
\end{equation}
In the dilute limit $c\to 0$, this reduces to
\begin{equation}
    N_b^0\equiv\lim_{c\rightarrow 0} N_b = \frac{M_m}{M_w}
    \frac{df(0)}{dc} \rho(0) = 
    \frac{r \rho_0}{\rho_0 + K_dM_w} 
\end{equation}
which can be recognized as the standard Langmuir absorption isotherm for 
water molecules at number density $\rho_0/M_w$ in the bulk at equilibrium 
with $r$ binding sites. 
If we neglect the correction due to the fraction of bound water
molecules that diffuse rigidly with dextran polymers,  
the first-order expansion of Eq.~\eqref{eq:f_solution} leads to
\begin{equation}
    \label{eq:Dnexp}
    \frac{D_{\rm obs}(c)}{D_0} \simeq 1 - \frac{c/M_m}{\rho_0/M_w} N_b^0 
    \simeq \exp\left(- \frac{c/M_m}{\rho_0/M_w} N_b^0\right)
\end{equation}
It can be recognized from Fig.~\ref{fig:fcDw}\subref{fig:fcDw-b}
that the exponential approximation is virtually indistinguishable from 
the full expression~\eqref{eq:f_solution}. The intriguing conclusion of this 
analysis is that the normalized diffusion coefficient of water in the presence
of dextran can be accurately described by a decreasing exponential of the 
number density of the crowders, $c/M_m$. The typical decay constant 
turns out to be the reduced bulk water number density
$\rho_0/(N_b^0M_w)$, which can be interpreted as bulk water number 
density per bound water molecule.  Interestingly, the best-fit values of $r$ and $K_d$ 
(see caption of Fig.~\ref{fig:fcDw}) give $N_b^0 = 19.5$, 
consistent with the average number of hydrogen bonds formed by a glucose molecule 
(the dextran monomer) with water molecules in the first hydration shell, which is around 
11~\cite{Suzuki2008}.

\subsection{Dextran self-diffusivities can be related to excess entropy scaling}
We examine next the relationship between exponential concentration dependence of the dextran self-diffusivities and excess entropy concepts. Introducing the polymer number concentration, 
\begin{equation}
   n = \frac{c N_A}{N M_m}
\end{equation}
and its value at $c = c^*$,
\begin{equation}
   n^* = \frac{c^* N_A}{N M_m} = \frac{3}{4 \pi R^3} \sim N^{-3\nu},
\end{equation}
we can express the reduced dextran self-diffusivity as
\begin{equation}
    \frac{D_{\rm DEX}}{D_{\rm DEX}^0} = e^{-\frac{n}{n^*}} \sim e^{- n R^3},
\end{equation}
we can recognize that the quantity $n R^3$ is proportional to the value of the second virial coefficient of a gas of hard spheres of radius $R$, $B_2 = \frac{16}{3} \pi R^3$ \cite{hill1960introduction}. 
Rosenfeld \cite{Rosenfeld1977} and Dzugutov \cite{Dzugutov2001} proposed that the self-diffusion coefficient of simple liquids can be written in the universal simple form \cite{Dyre2018}
\begin{equation}
    \label{eq:ex_ent_scal}
    D = {n^{\frac{1}{3}}\sqrt{\frac{m}{kT}}}e^{-\alpha s_{ex}},
\end{equation}
where $s_{ex}$ is the excess entropy per particle, $m$ is the particle mass and $T$ is the temperature of the system. In the excess entropy scaling approach, the mass of the particle is a fixed parameter and therefore the pre-factor of the exponential in eq.~\eqref{eq:ex_ent_scal} scales as $n^{\frac{1}{3}}\sim {l_0}^{-1}$, where $l_0$ is the characteristic length of the system \cite{Dyre2018}. 
The pre-factor of the scale-invariant exponential in eq.~\eqref{eq:scaleddiff}, $D_0$,  scales as the inverse of the radius of gyration (in dilute solution): $D_0 \sim R^{-1}$. Indeed $R$ is the fundamental length scale of polymer solutions, determining the overlap concentration. 
In the two-body approximation, the excess entropy reads \cite{Dyre2018, Korkmaz2006}
\begin{equation}
    \label{eq:s_ex}
    s_{ex} \simeq s_2 = 2\pi n\int_0^{\infty} \left[g(r)\ln\left(g(r)\right) - g(r) + 1 \right]r^2dr
\end{equation}
where $g(r)$ is the intermolecular radial distribution function of polymers and $r$ is the centre-of-mass distance between the polymers chains. It is well known that for hard spheres, the integral~\eqref{eq:s_ex} gives the volume that a sphere excludes to the center of mass of other spheres. Therefore one has
\begin{equation}
    \label{eq:s_2}
    s_{2} = \frac{16}{3} \pi n R^3  = 4\frac{n}{n^*} 
\end{equation}
The simplest way to model partially overlapping spheres beyond $n/n^* = 1$ is the so-called cherry-pit model \cite{torquato2002random}, where the spherical particles have a hard core of size $\lambda R < R$ and can freely overlap for distances greater than $2 \lambda R$. If $\lambda$ is small and independent from $N$, then the two-particle excess entropy becomes
\begin{equation}
    \label{eq:s2cherry_pit}
        s_{2} = \frac{16}{3} \pi n \lambda^3 R^3 = 4 \lambda^3\frac{n}{n^*}  = 4\lambda^3\frac{c}{c^*}.
\end{equation}
We conclude that for partially overlapping spheres $n/n^*$ may reach values as high as $s_2/4\lambda^3$ .

According to Dzugotov's definition of excess entropy scaling\cite{Dzugutov2001}, $\alpha = 1$ in relation~\eqref{eq:ex_ent_scal}. Hence, letting $c^* = \overline{c_1}N^{1-3\nu}$, combining Eq.~\eqref{eq:scaleddiff} and Eq.~\eqref{eq:s2cherry_pit}), an effective size of the hard-core of the crowders can be defined, namely
\begin{equation}
    \overline{c_1} = 4 \lambda^3 c_1 
\end{equation}
In the hard-sphere case ($\lambda$ = 1),  $\overline{c_1} = 4c_1 = 2.4 \, {\rm g/cm^3}$. Therefore, an independent estimate of $\overline{c_1}$ from the data provides a direct measure of the effective core size $\lambda$.According to its definition ~\eqref{cond:c_overlap}, the overlap concentration can be expressed in units of mass concentration as
\begin{equation}
    \label{eq:c1def}
    c^* = \frac{M_m N^{(1-3\nu)}}{N_A\left(\frac{4\pi}{3}a^3\right)}\equiv \overline{c_1} N^{(1 - 3\nu)}.
\end{equation}
Here, $a$ is the radius of the monomer if it were a sphere. According to this definition, with the hard-sphere estimation $\overline{c_1}$ = 2.4 g/cm$^3$, $a$ = 0.30 nm, a value that captures the order of magnitude of the size of dextran monomer. If $a$ is identified with the effective hydrodynamic radius of D-glucose, $a$ $\sim$ 0.45 nm \cite{Schultz1961}, Eq.~\eqref{eq:c1def} yields $\overline{c_1}$ = 0.70 g/cm$^3$ and therefore:
\begin{equation}
    \lambda = \left(\frac{\overline{c_1}}{4c_1}\right)^{\frac{1}{3}} \sim 0.66
\end{equation}
We note that the hard sphere interaction is not a necessary condition for the condition $s_{2} \sim R^3$ to hold. Indeed, excess entropy scaling applies also to systems with soft sphere interactions \cite{Rosenfeld1977}. If the polymer radial distribution function is an invariant function of the polymer radius in dilute conditions $g(r) \sim g\left(\frac{r}{R}\right)$, then the integral in Eq.~\eqref{eq:s_ex} gives a contribution proportional to $R^3$ and the condition $s_{2} \sim R^3$ holds. 
One can approximate $g(r)$ with the two-particle distribution function
\begin{equation}
    g(r) \sim g_0(r) = e^{-\frac{u(r)}{kT}},
\end{equation}
where $g(r)$ is an invariant function of $r/R$:
\begin{equation}
    \label{cond:g(r)}
    g(r) \sim \tilde{g} (\frac{r}{R}) \equiv g(r').
\end{equation}
Then, we have
\begin{equation}
    s_2 = 2\pi n R^3\int_0^{\infty} \left[g(r')\ln\left(g(r')\right) - g(r') + 1 \right]r'^2dr'\sim  n R^3 I(n).
\end{equation}
That is, $s_2$ would result in the hard sphere expression, multiplied by an integral $I(n)$, that could only depend on the concentration; the condition in Eq.~\eqref{cond:g(r)} eliminates any dependence from $N$ in $I(n)$. If we identify $u(r)$ with the interaction potential between two polymers, it was argued that such an interaction must be independent from $N$ for linear polymers with the excluded volume interaction (SAWs) \cite{Grosbergetal1982}. If this applied to dextran, it would confirm our hypothesis for $g(r)$. In the case of linear self-avoiding polymers, this hypothesis was investigated through simulations~\cite{Bolhuisetal2001}, which showed that the polymer-polymer interaction for SAWs could be considered as soft and it was found almost invariant with concentration, up to $n/n^* \sim 8$. If this property applied to dextran polymers, it would justify the required scaling properties. The breakdown for DEX450 at higher concentrations may be related to the breakdown of this approximate regime. 

With excess entropy scaling, typically the universal behavior is studied as a function of particle concentration and temperature. In our case, the temperature was fixed, and it is the degree of polymerization that plays the role of the variable coupled to concentration. Even though we find it useful to surmise a connection between the typical scaling laws of polymer physics and the excess entropy scaling hypothesis, we cannot rigorously state that the system satisfies excess entropy scaling, at least for the following two reasons: first, other temperatures should be explored, and second, the excess entropy of the system should be quantified independently. 

Excess entropy scaling typically applies also, even though to a lesser extent \cite{Rosenfeld1977, Dyre2018}, to viscosity. In our case, the asymptotic trends of the concentration-dependent viscosity are clearly power-law and we could not fit the data with an analogous exponential function. We have also seen that the Stokes-Einstein product $D_{\rm DEX}.\eta$ is not constant but increases with concentration. 

\subsection{The Flory exponent of dextran signifies a branched polymer}

The low value of the Flory exponent $\nu \sim 0.44$ is typically interpreted as a sign of branching of dextran \cite{Nordmeier1993,Bu:Art,Antoniou2012} and the question whether dextran could be attributed to either hyperbranched or randomly branched universality classes has already been investigated, although not conclusively to our knowledge \cite{IoAbBu:Art,IoAbBu_3_2001}. 

Interestingly, the value $\nu \sim 0.44$, obtained in this study independently from viscosity and dextran self-diffusion measurements, is very close to 
the value $\nu$ = 7/16 predicted by the Flory theory of quenched randomly branched polymers with ideal connectivity in theta solvent \cite{EvGrRuRo:Art, daoud1981conformation}. A Flory exponent $\nu < 0.5$ has also been found in a Monte Carlo simulation \cite{CuCh:Art} for branched quenched polymers in good solvents, while other simulations provide best estimates lower than, but compatible with, 1/2 \cite{redner1979mean,CuCh:Art}. However, the value predicted by Flory theory and renormalization group calculations is exactly $\nu$ = 1/2 for branched quenched polymers in good solvent\cite{EvGrRuRo:Art,isaacson1980flory,Parisi1981}. The possible identification of dextran with a well-defined class of branched polymers is still an open question. 
We cannot rule out other explanations for $\nu \sim 0.44$. First of all, to conclusively investigate the topological features of dextran, polydispersity, which is expected to increase with branching, should be considered. Coupled with a possible dependence of the branching properties from $N$, polydispersity may affect the value of $\nu$. However, Burchard and coworkers, explicitly addressed polydispersity of dextran and found it increased slowly from a value of $M_{\rm weight}/M_{\rm number}$ $\sim$ 2 for small molecular weights, before increasing significantly towards a power law regime at molecular weights greater than $2\times10^6$ g/mol\cite{Ioan2000}. Their results suggest that polydispersity of DEX10 and DEX40 should be well approximated by $M_{\rm weight}/M_{\rm number}$ $\sim$ 2, while in the range of DEX450 one may expect values around $M_{\rm weight}/M_{\rm number}$ $\sim$ 3. As long as its variation is confined in this range, polydispersity alone cannot explain the low value of the Flory exponent. Indeed, values of $\nu$ lower than 0.5 are reported for both monodisperse and polydisperse dextran samples \cite{venturoli2005ficoll}. 
In the theory of randomly branched polymers, the degree of branching is assumed to be independent of the degree of polymerization, $N$. If dextran polymers were almost linear for low $N$ and the probability of branching per monomer increased with $N$, then the scaling behavior would be just illusory. Burchard and coworkers found an evident deviation from power law behavior for intrinsic viscosity \cite{Ioan2000, IoAbBu_3_2001}. Nevertheless, direct estimates of the radius of gyration of dextran in dilute conditions are generally well described by power laws\cite{Ioan2000}, suggesting that the behavior of viscosity may be influenced by other factors, without doubting the scaling property of the radius of gyration. \\
The value 7/16 for the Flory theory of quenched branched polymers should be realized at a single $\theta$-temperature. However, experimentally the value of $\nu$ $\sim$ 0.44 is quite independent from temperature \cite{Nordmeier1993}. An estimation of $\theta$-temperature around 43°C for dextran was given by Guner \cite{Guner1999}. In the same work, however, an exponent a = 0.513 in the Mark-Houwink relation at 37°C was estimated, which would give rise to $\nu \simeq 0.504$, at odds with the previously cited results. 

Another fact that may be considered is aggregation: Burchard and collaborators attributed the breakdown of universal behavior of different dextran sizes in semidilute solutions to the association of polymers \cite{IoAbBu:Art}. We notice that for clusters of polymers the interaction range would be necessarily greater than $2R$, leading to a natural breakdown of the hypotheses at the basis of excess entropy scaling.

\subsection{Three measures of the volume fraction}
Estimating volume fractions of hard spheres is simple. The main point in this section is that estimating volume fractions of hydrated macromolecules such as dextran is not as clean as one would wish. 
Previous work on Ficoll solutions~\cite{Ranganathan2022} has identified the important contribution of bound water on hydrated crowders in reducing the free volume significantly. Following this, we introduce three types of measures of volume concentrations. The fraction volume of solute $\phi_s(c)$ can be estimated assigning a specific volume $\overline{v}$ to each mass unit of solute: 
\begin{equation}
    \label{eq:phi_s}
    \phi_s(c) = \overline{v}c
\end{equation}
However, we have seen that dextran in water gets highly hydrated.
We may calculate, based on the water diffusion measurements, the volume fraction based on the volume of hydrated dextran:
\begin{equation}
    \label{eq:phi_h}
    \phi_h (c) =  f(c) + c\overline{v}\left(1 - f(c)\right).
\end{equation}
Finally, we have  seen in the previous section that $c/c^*$ can be interpreted as a macromolecular packing fraction, i.e. the packing fraction of dextran macromolecules if one wants to consider them as compact objects, such as overlapping spheres. Shown in Figure \ref{fig:vol_conc} are the overall results (details in the Appendix).
%
 \begin{figure}[t!]
  \includegraphics[width=\textwidth]{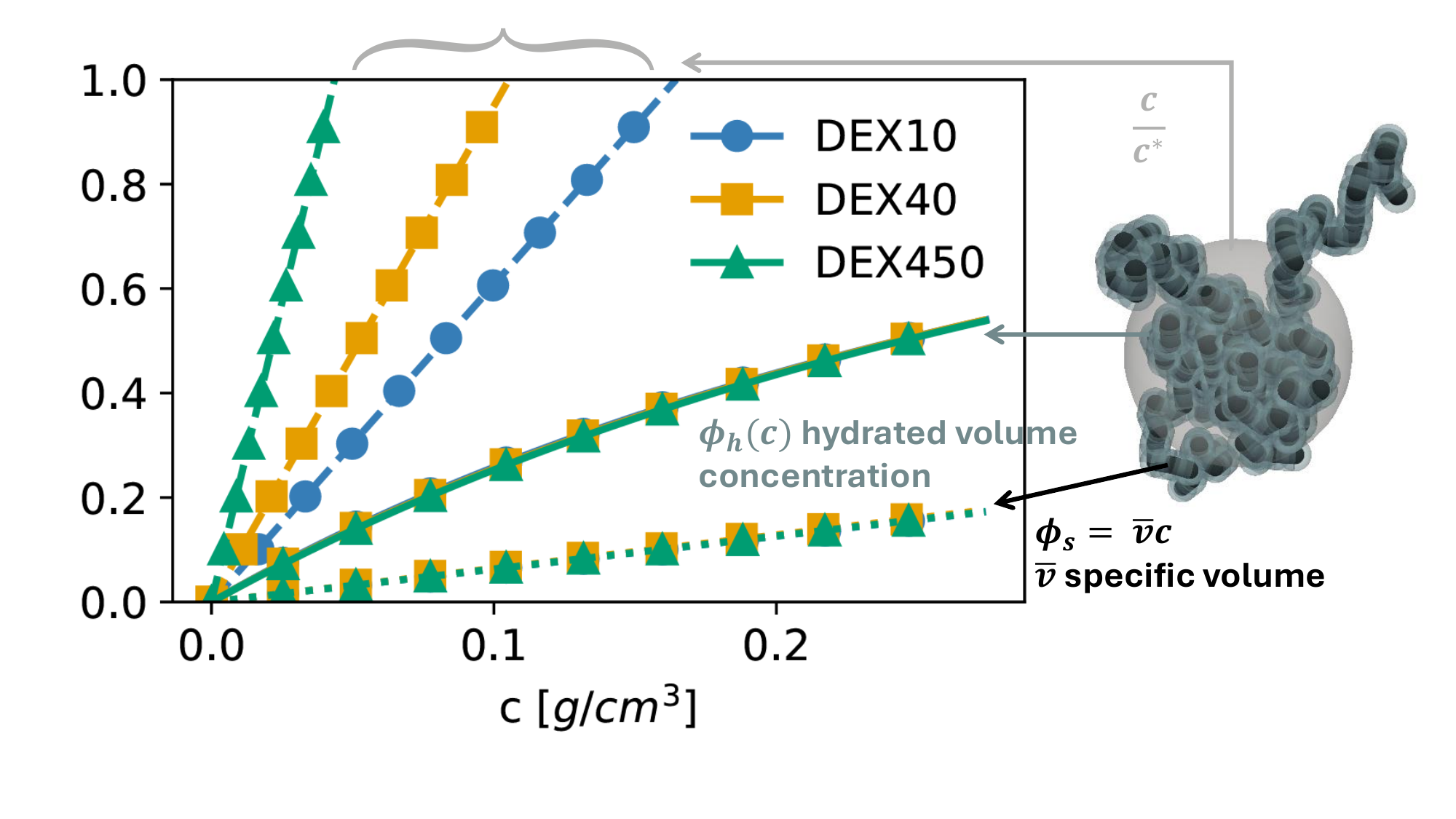}
   \caption{Comparison of three distinct measures of volume concentration. $\phi_s (c)$, eq. ~\eqref{eq:phi_s}, dotted line, corresponds to the black core volume in the 3D polymer sketch; $\phi_h (c)$, eq. ~\eqref{eq:phi_s}, continuous line, is associated with the slate gray-shaded hydrated volume; $c/c^*$, dashed line, represents the light gray-shaded sphere, with radius equal to the radius of gyration. The hydrated $\phi_h (c)$ volume fraction is systematically greater than the dry one ($\phi_s (c)$), by a factor of some units. The macromolecular packing fractions $c/c^*$ (dashed lines), however, are significantly greater and reach 1 in the experimental range.  At odds with $\phi_s (c)$ and $\phi_h (c)$, the ratio $c/c^*$ is obviously size dependent. Here the overlap concentration $c^* = c_1N^{1-3\nu}$ is estimated for the three sizes according to the parameters of the cumulative self-diffusion fit (eq. \ref{eq:scaleddiff}), with $N = M_{\rm DEX}/M_m$. If for the overlap concentrations the estimates from the viscosity were used, the initial slopes would even be greater.} 
   \label{fig:vol_conc}
\end{figure}
%
As expected, the hydrated fraction $\phi_h(c)$ is greater than the dry one, $\phi_s(c)$, over the whole experimental range. However, $c/c^*$ increases much more rapidly with concentration and the overlap concentration is reached early. 
Often, in macromolecular crowding studies, polymer crowders are treated as hard spheres or ellipsoids\cite{ZimmermanMinton1993,Stagg2007}, whose volume is determined by the specific volume\cite{Homchaudhuri2006}. For hard spheres, the scaling exponent is $\nu$ = 1/3, and therefore $c^*$ does not show any size dependence. When modeling a polymeric crowder, it is important to distinguish between the different measures of packing fractions. Hard or soft sphere models may be used to describe the interaction between crowders and other macromolecules, because viscosity, self-diffusion and thermodynamic quantities such as the virial coefficients, if the concentration $c$ is not too high, are known to scale with the overlap concentration $c^*$\cite{IoAbBu:Art,Teraoka2002}. In other words, their scaling properties are controlled by the dilute radius $R \sim N^\nu$. If the physical observable under scrutiny in crowded conditions depends explicitly on the interaction with the solvent, $\phi_s(c)$ or $\phi_h(c)$ are more suitable measures, as long as they scale linearly with the total number of monomers in solution. 
Finally, this reasoning suggests to perform macromolecular crowding experiments in good (or theta) solvents, with polymer crowders of similar chemistry but different topology, i.e. linear or branched.

\section{Conclusions}

In this work, several physical properties of dextran solutions were studied experimentally and discussed in the context of polymer scaling. The relative viscosity of the solution shows a classic crossover from dilute to semi-dilute polymer behavior. Scaling concentrations by the intrinsic viscosity for three polymer molecular weights, the viscosity of all solutions collapsed onto a single universal curve. 

The self-diffusivity of dextran was experimentally probed using PFG-NMR. The trend could be accounted very well by exponential decays, especially for DEX10 and DEX40. A cumulative fitting for a reduced self-diffusion variable can globally account for the observed trends, scaling by $c^*_D$, a quantity that can be considered as a proxy for the overlap concentration, $c^*$. Only for DEX450 at high concentration (in the semi-dilute regime), the data were found to deviate from the universal exponential trend. We attempted to connect the universal behavior of the reduced self-diffusion with excess entropy scaling. Focusing only on the two-particle entropy $s_2$, it is shown how, in not too concentrated conditions, the dilute radius emerges as the only relevant length scale of the solution. This leads directly to the two-particle excess entropy $s_2$ being directly proportional to $c/c_D^*$. 
It might be possible to extend this reasoning to other polymers different from dextran and characterized by a different topology. This interpretation is rooted in the idea that dextran self-diffusion, as well as viscosity, are driven by macromolecular inter-polymer interactions. 

The distinct universal behaviors found for viscosity and the reduced polymer self-diffusion provide us with two independent estimations of the Flory exponent for dextran that are in agreement with each other: $\nu$ $\sim $ 0.44. This low value for the Flory exponent is recurrent in the literature of dextran and is typically related to branching. Previous studies attempted to associate dextran either to the class of hyperbranched or randomly branched  polymers\cite{Nordmeier1993,Bu:Art,IoAbBu_3_2001}. Although in this work we cannot draw a definitive conclusion, we note that  $\nu$ $\sim$ 0.44 is the expected exponent for the Flory theory of quenched randomly branched polymers in $\theta$ - solvent\cite{daoud1981conformation,EvGrRuRo:Art}. A definitive assessment of the topological properties of dextran would be extremely useful for its characterization as a crowding agent.

At odds with viscosity and self-diffusion, the density of dextran solutions and the diffusion of water molecules do not show any dependence on the size of dextrans as a function of mass concentration. We interpret this result as a clear sign that these properties are driven by interactions between the solvent and the monomers. A model admitting two states, either bound or free in solution, for water and dextran monomers is demonstrated to account for the decreasing trend of water diffusion. 

The measurement of the diffusion of water also provides an estimate of the total hydrodynamic volume of dextran in solution. This latter, when viewed in comparison with the more usual volume fraction estimated by dextran specific volume, shows how important can be to consider the contribution of bound water in the estimation of available volume. However, these two estimations of volume concentration are significantly lower than $c/c^*$, which unlike $\phi_s(c)$ or $\phi_h(c)$, exhibits a dependence on dextran molecular weight. 
These observations highlight the importance of what has already been noted in other contexts, i.e., that determinations of excluded volume are known to depend on the tracer dimensions\cite{Minton2001}, so that the scale of the (macro)molecule of interest determines the degree of crowding it experiences.

This work suggests that, when using polymer crowders, physical quantities driven by inter-molecular interactions in crowded conditions may generally show a dependence on the degree of polymerization, in a way that can be interpreted in terms of scaling concepts from polymer physics. On the other hand, physical quantities dominated by the monomer-solvent interaction are not expected to depend on polymer size. Systematically varying the size of the crowder in crowding experiments serves to highlight this difference.

Finally, we have presented a quantitative examination of the polymer physics of dextran crowders. One hope underlying this work is that quantitative studies in macromolecular crowding will use similarly detailed studies of the crowder prior to reporting on the effects of crowding itself.

\section{Acknowledgments}
The NMR technical support of C\'eline Schneider (MUN), Herv\'e Meudal and Karine Loth (CBM) is gratefully acknowledged. This work was supported by the French Agence Nationale de la Recherche (ANR), through the project X-CROWD, contract ANR-21-CE44-0020-01.
A.Y. acknowledges the financial support of the National Science and Engineering Research Council of Canada (RGPIN-2025-05494). 
\bibliography{refs,bibliography_GM}

\newpage
\section{Appendix: Volume fraction calculations}
\label{sec:Vol_frac_calc}
Through the water bound fraction $f(c)$, one can estimate a volume concentration $\phi_h$ that accounts also for the hydration of dextran. Indeed $\phi_h$ can be identified with the fraction of the total solution volume $V$ that cannot be attributed to free solvent 
\begin{equation}
    \label{eq:phi_h_0}
    \phi_h = \frac{V - V_{free}}{V}
\end{equation}

\noindent The total mass of water in solution is $M(1-w) = M(1-c/\rho(c))$. Therefore, one can estimate $V_{free}$ through the bound water fraction $f(c)$ and the solvent density $\rho_0$
\begin{equation}
    \label{eq:V_free}
    V_{free} = \frac{M}{\rho_0}\left(1 - c/\rho(c)\right)\left(1 - f(c)\right)
\end{equation}
Substituting eq.~\eqref{eq:V_free} in eq.~\eqref{eq:phi_h_0}, the latter reads
\begin{equation}
    \phi_h (c) = 1 - \frac{1}{\rho_0}\left(\rho(c) - c\right)\left(1 - f(c)\right)
\end{equation}
which, using eq.~\eqref{eq:density_c}, 
$\rho(c)=\rho_0 + c(1-\rho_0\bar{v})$, can be expressed as :
\begin{equation}
    \phi_h (c) =  f(c) + c\overline{v}\left(1 - f(c)\right)
\end{equation}
The initial slope at low concentration is:
\begin{equation}
    \frac{d\phi_h}{dc} (0) = \frac{df}{dc}(0) + \overline{v}
\end{equation}
It can be checked (eq.~\eqref{eq:f_solution}) that $\frac{df}{dc}(0) \approx$ 2.2 cm$^3$/g. Therefore, according to this estimation and considering that $\overline{v} \approx$ 0.63 cm$^3$/g, the hydration water accounts for roughly $2.2/(2.2+0.63) \approx 77$ \% of the hydrated volume fraction $\phi_h$ 
at low concentration.

\end{document}